\documentclass[reprint,prl,preprintnumbers,superscriptaddress,amsmath,amssymb]{revtex4-1}
\usepackage{hyperref}
\usepackage{float}
\usepackage{natbib}
\usepackage{graphicx}
\usepackage{dcolumn}
\usepackage{bm}
\usepackage{siunitx} 

\begin{document}

\bibliographystyle{unsrtnat}
\title{\large {Valley-selective optical Stark effect in monolayer WS$_2$}}

\vspace{0.5cm}
\author{Edbert J. Sie}
		\affiliation{Department of Physics, Massachusetts Institute of Technology, Cambridge, MA 02139, USA}
\author{James W. McIver}
		\affiliation{Department of Physics, Massachusetts Institute of Technology, Cambridge, MA 02139, USA}		
		\affiliation{Department of Physics, Harvard University, Cambridge, MA 02138, USA}
\author{Yi-Hsien Lee}
		\affiliation{Material Sciences and Engineering, National Tsing-Hua University, Hsinchu 30013, Taiwan}
\author{Liang Fu}
		\affiliation{Department of Physics, Massachusetts Institute of Technology, Cambridge, MA 02139, USA}
\author{Jing Kong}
		\affiliation{Department of Electrical Engineering and Computer Science, Massachusetts Institute of Technology, Cambridge, MA 02139, USA}		
\author{Nuh Gedik}
		\altaffiliation{gedik@mit.edu}
		\affiliation{Department of Physics, Massachusetts Institute of Technology, Cambridge, MA 02139, USA}
\date{\today}
\vspace{0.5cm}

\maketitle

\noindent
\textbf{\small{Breaking space-time symmetries in two-dimensional crystals (2D) can dramatically influence their macroscopic electronic properties. Monolayer transition-metal dichalcogenides (TMDs) are prime examples where the intrinsically broken crystal inversion symmetry permits the generation of valley-selective electron populations \cite{DiXiao12,KFMak12,HZeng12,TCao12}, even though the two valleys are energetically degenerate, locked by time-reversal symmetry. Lifting the valley degeneracy in these materials is of great interest because it would allow for valley-specific band engineering and offer additional control in valleytronic applications. While applying a magnetic field should in principle accomplish this task, experiments to date have observed no valley-selective energy level shifts in fields accessible in the laboratory. Here we show the first direct evidence of lifted valley degeneracy in the monolayer TMD WS$_2$. By applying intense circularly polarized light, which breaks time-reversal symmetry, we demonstrate that the exciton level in each valley can be selectively tuned by as much as 18 meV via the optical Stark effect. These results offer a novel way to control valley degree of freedom, and may provide a means to realize new Floquet topological phases \cite{Inoue10,Kitagawa11,Lindner11} in 2D TMDs.}}

The coherent interaction between light and matter offers a means to modify and control the energy level spectrum of a given electronic system. This interaction can be understood using what is known as Floquet theory \cite{Shirley65}, which states that a Hamiltonian periodic in time has quasistatic eigenstates that are evenly spaced in units of the photon driving energy. The simplest example of this is given by a two-level atomic system in the presence of monochromatic light, which can be fully described by the semi-classical Hamiltonian
\begin{equation}
\hat{H}(t) = \hat{H}_0 + \hat{p}\mathcal{E}(t)
\end{equation}
where $\hat{H}_0$ is the equilibrium Hamiltonian describing a two-level atom with eigenstates $\left|a\right\rangle$ and $\left|b\right\rangle$, $\hat{p}$ is the electric dipole moment operator of the atom, and $\mathcal{E}(t) = \mathcal{E}_0\cos 2\pi\nu t$ is the oscillating electric field of light with amplitude $\mathcal{E}_0$ and frequency $\nu$. The perturbation term in the Hamiltonian contains a time-dependent factor ($\cos 2\pi\nu t$) that enables the coherent absorption of light by $\left|a\right\rangle$ to form the photon-dressed state $\left|a+h\nu\right\rangle$, and the stimulated emission of light by $\left|b\right\rangle$ to form another photon-dressed state $\left|b-h\nu\right\rangle$, as shown in Fig 1a. Additional states $\left|a-h\nu\right\rangle$ and $\left|b+h\nu\right\rangle$, as well as higher order terms, also arise in this situation, but we omit them from Fig 1a for the sake of clarity. This semi-classical description is consistent with the fully quantized approach (Supplementary).
\begin{figure*}[t]
	\includegraphics[width=0.75\textwidth]{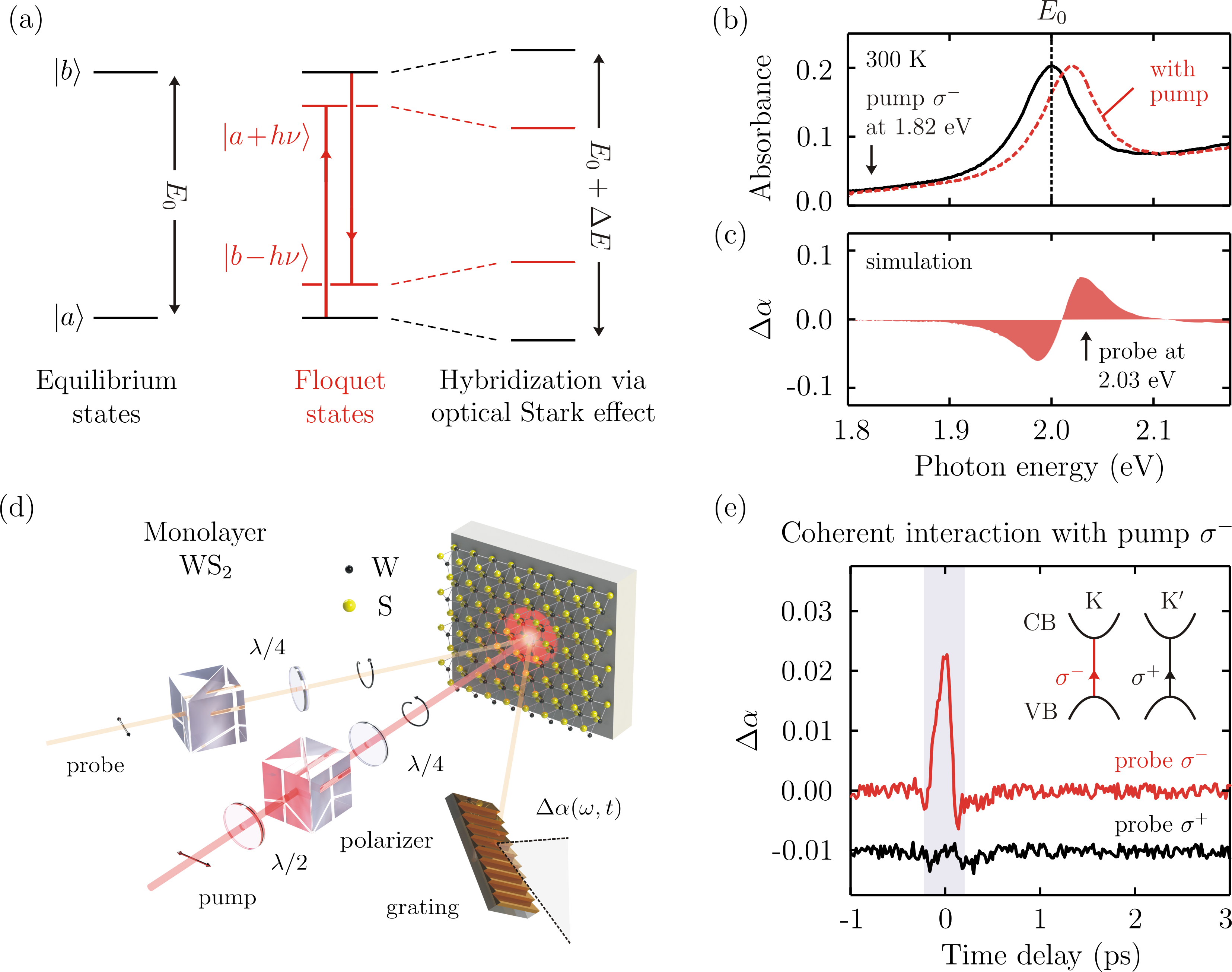}
	\caption{\textbf{The optical Stark effect and its observation in monolayer WS$_2$.} (a) Energy level diagram of two-level atoms showing that the equilibrium and Floquet states can hybridize, resulting in shifted energy levels. (b) Measured absorbance of monolayer WS$_2$ (black) and a rigidly shifted one (dashed) to simulate the optical Stark effect. (c) The simulated change of absorption induced by the pump pulses. (d) Schematic of transient absorption spectroscopy setup (see Methods). (e) Time-trace of $\Delta\alpha$, induced by pump pulses of $\sigma^-$ helicity, measured using probe pulses of the same ($\sigma^-$, red) and opposite ($\sigma^+$, black, with offset for clarity) helicities. It shows that the optical Stark effect only occurs during the pump pulse duration ($\Delta t =$ 0 ps) and when the pump and probe helicities are the same.}
	\label{fig:Fig1}
\end{figure*}

The series of photon-dressed states formed in this way are called the Floquet states, which can hybridize with the equilibrium states $\left|a\right\rangle$ and $\left|b\right\rangle$ through the electric field term $\mathcal{E}$ in equation (1). This hybridization often results in energy repulsion between the equilibrium and Floquet states, the physics of which is equivalent to the hybridization between two atomic orbitals by the Coulomb interaction to form the bonding and anti-bonding molecular orbitals in quantum chemistry. The interaction between the two states always results in a wider energy level separation, and the magnitude of the energy repulsion becomes more substantial if the states are energetically closer to each other. In our case, there are two such pairs of states shown in Fig 1a denoted as $\left\{\left|a+h\nu\right\rangle,\left|b\right\rangle\right\}$ and $\left\{\left|a\right\rangle,\left|b-h\nu\right\rangle\right\}$ with identical energy difference of $\Delta = (E_b - E_a) - h\nu$. Through the simultaneous energy repulsion of these pairs, the optical transition between states $\left|a\right\rangle$ and $\left|b\right\rangle$ is shifted to a larger energy. This phenomenon is known as the optical Stark effect \cite{Autler55, Bakos77}, and the corresponding energy shift ($\Delta E$) is given by
\begin{equation}
\Delta E = \frac{\mathcal{M}_{ab}^2 \left<\mathcal{E}^2\right>}{\Delta}
\end{equation}
where $\mathcal{M}_{ab}$ is the polarization matrix element between $\left|a\right\rangle$ and $\left|b\right\rangle$, and $\left<\mathcal{E}^2\right>$ ($= \mathcal{E}_0^2/2$) is the time-averaged value of the electric field squared, proportional to the light intensity (Supplementary). Owing to the tunability of this energy shift by changing either the light intensity or frequency, the optical Stark effect has been routinely employed in the study of atomic physics and to facilitate the atomic cooling below the Doppler limit \cite{Cohen90}.

Similar effects have been encountered in solid-state systems, where the electronic states are in the form of Bloch bands that are periodic in momentum space. In the presence of monochromatic light, the Hamiltonian of crystalline solids becomes periodic in both space and time, which leads to the creation of Floquet-Bloch bands that repeat in momentum and energy. Floquet-Bloch bands were very recently observed for the first time on the surface of a topological insulator irradiated by mid-infrared light \cite{YHWang13}. The optical Stark effect can also occur in solids through the interaction between photo-induced Floquet-Bloch bands and equilibrium Bloch bands \cite{Joffre88}. To date, this effect has only been reported in a very limited number of materials, with Cu$_2$O \cite{Frohlich85}, GaAs \cite{Mysyrowicz86, Chemla89, Sieh99, Hayat12} and Ge \cite{Koster11} semiconductors among the few examples.

Here we report the first observation of the optical Stark effect in a monolayer TMD WS$_2$. The recently discovered monolayer TMDs are 2D crystalline semiconductors with unique spin-valley properties \cite{DiXiao12} and energetically degenerate valleys at the K and K$^{\prime}$ points in the Brillouin zone that are protected by time-reversal symmetry. We demonstrate by using circularly polarized light that the effect can be used to break the valley degeneracy and raise the exciton level at one valley by as much as 18 meV in a controllable valley-selective manner.
\begin{figure*}[t]
	\includegraphics[width=0.97\textwidth]{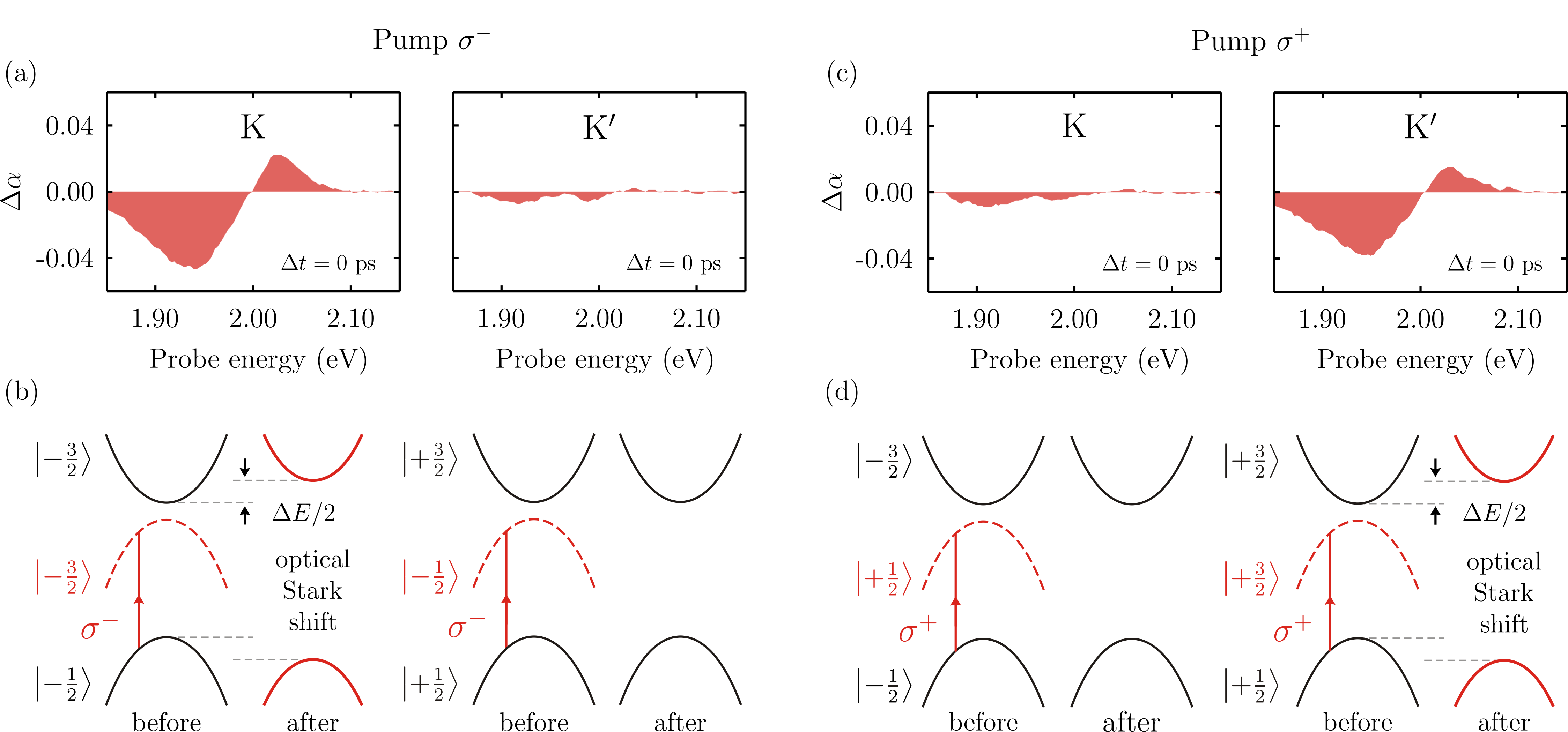}
	\caption{\textbf{The valley-selectivity of the optical Stark effect.} (a) Valley-specific $\Delta\alpha$ spectra (BG-subtracted) induced by $\sigma^-$ pump pulses probed by using $\sigma^-$ (K valley) and $\sigma^+$ (K$^{\prime}$ valley) helicities. (b) Band diagrams with the pump-induced Floquet-Bloch bands (red dashed curves) and their associated magnetic quantum numbers $m$. By switching the pump pulse helicity to $\sigma^+$, we also measured the valley-specific $\Delta\alpha$ spectra in (c) and shown their corresponding band diagrams in (d). It shows that the optical Stark effect only occurs via states having the same magnetic quantum numbers.}
	\label{fig:Fig2}
\end{figure*} 

In monolayer WS$_2$, the energy of the lowest exciton state is 2.00 eV at room temperature (Fig 1b, black). In order to induce a sufficiently large energy shift of this exciton (as simulated and exaggerated for clarity in Fig 1b, dashed), we use an optical parametric amplifier capable of generating ultrafast laser pulses with high peak intensity and tunable photon energy (1.68$-$1.88 eV). To measure the energy shift we use transient absorption spectroscopy (Fig 1d), which is a powerful technique capable of probing the resulting change in the absorption spectrum $\Delta\alpha$, as simulated in Fig 1c \cite{EJSie13} (see Methods). The unique optical selection rules of monolayer TMDs allow for $\Delta\alpha$ at the K (or K$^{\prime}$) valley to be measured with valley-specificity by using left (or right) circularly polarized probe light (inset in Fig 1e).

To search for the optical Stark effect in WS$_2$, we start by measuring the change in the optical absorption $\Delta\alpha$ as a function of time delay $\Delta t$ between the pump and probe laser pulses (Fig 1e). To generate the necessary Floquet-Bloch bands, we tune the pump photon energy to 1.82 eV so that it is just below the absorption peak, with pulse duration 250 fs at FWHM, fluence \SI{60}{\micro\J}/cm$^2$, and polarization $\sigma^-$ (left circularly polarized). Since the optical Stark effect is expected to shift the absorption peak to higher energy, the probe photon energy is chosen to be 2.03 eV, which is above the equilibrium absorption peak. Fig 1e shows that when $\sigma^-$ is used to probe $\Delta\alpha$ (red trace), there is a sharp peak that only exists at $\Delta t = 0$ ps when the pump pulse is present. This signifies that we are sensitive to a coherent light-matter interaction occurring between the pump pulse and the sample. When $\sigma^+$ is used to probe $\Delta\alpha$ (black trace), we observe no discernible signal above the noise level at all time delays. This shows that probing the optical Stark effect in this material is strongly sensitive to the selection of pump and probe helicities. Closer examination of $\Delta\alpha$ spectrum in the range of 1.85$-$2.15 eV (Supplementary) reveals a faint but noticeable background signal that is present at both valleys. In the following discussion, we only consider results taken at $\Delta t = 0$ ps for which the background signals (BG) have also been subtracted in order to focus on the optical Stark effect (see Supplementary).
\begin{figure*}[t]
	\includegraphics[width=0.70\textwidth]{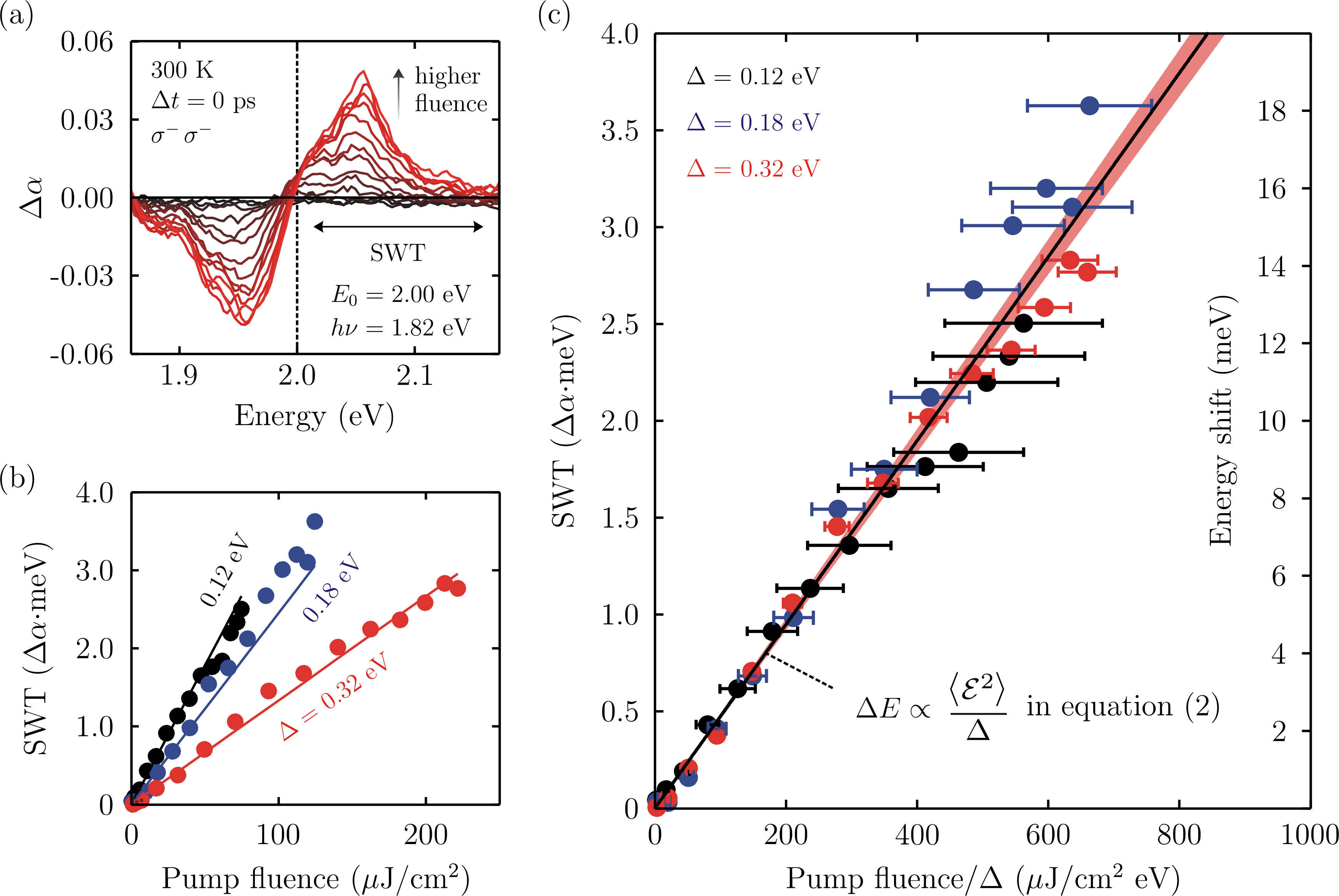}
	\caption{\textbf{Fluence and detuning dependences of the optical Stark shift.} (a) Fluence dependence of $\Delta\alpha$ spectra with fluences up to \SI{120}{\micro\J}/cm$^2$ measured with the same pump and probe helicities. (b) Fluence and detuning dependences of the spectral weight transfer (SWT) with integration range shown in (a). (c) SWT plotted as a function of fluence/$\Delta$, showing that all the data points fall along a common slope (black line). The horizontal error bars correspond to the pump bandwidth of 43 meV at FWHM. The fitting slope (black line) and the 95\% confidence band (red shades) shows an excellent agreement with the characteristic dependences of the optical Stark shift in equation (2).}
	\label{fig:Fig3}
\end{figure*}

Having demonstrated that the optical Stark effect can be induced in a monolayer TMD, we now study its valley-specificity and demonstrate that it can be used to lift the valley degeneracy. Fig 2a shows the $\Delta\alpha$ spectra induced by $\sigma^-$ pump pulses probed in a valley-selective manner by using $\sigma^-$ (K valley) and $\sigma^+$ (K$^{\prime}$ valley) probe helicities. When the same pump and probe helicities are used ($\sigma^-\sigma^-$, Fig 2a left panel), the spectrum displays a positive (and negative) $\Delta\alpha$ at energy higher (and lower) than the original absorption peak $E_0 (=2.00$ eV), clearly indicates that the absorption peak at the K valley is shifted to higher energy via the optical Stark effect. The spectrum of the opposite pump and probe helicities ($\sigma^-\sigma^+$, Fig 2a right panel), in contrast, shows a negligible signal across the whole spectrum, indicating an insignificant change to the absorption peak at the K$^{\prime}$ valley. Fig 2c shows identical measurements of $\Delta\alpha$ using instead a $\sigma^+$ pump helicity, where it can be seen that the effect is switched to the K$^{\prime}$ valley. This valley-selective probing of the optical Stark shift shows that the effect is well isolated within a particular valley determined by the pump helicity. The magnitude of the effect in either valley can also be smoothly tuned as the pump helicity is continuously varied from fully $\sigma^-$ to fully $\sigma^+$ (Supplementary).

The valley-specific optical Stark effect shares the same origin with the valley-selective photoexcitation as previously reported in this class of materials \cite{KFMak12, HZeng12, TCao12}. Both effects arise due to valley selection rules. In monolayer WS$_2$, the highest-occupied states in the valence band (VB) are associated with magnetic quantum numbers $m$, where $m = -1/2$ and $+1/2$ at K and K$^{\prime}$ valleys, respectively, as shown in Fig 2b and repeated in Fig 2d. Meanwhile, the lowest-unoccupied states in the conduction band (CB) consist of four quasi-degenerate states, with $m = -3/2$ and $-1/2$ at K valley, and $m = +3/2$ and $+1/2$ at K$^{\prime}$ valley. Two of these states ($m = \pm 1/2$) have no role in the effect, thus omitted from the figures. Coherent absorption of light by the VB creates a Floquet-Bloch band $\left|\text{VB}+h\nu\right\rangle$ for which the magnetic quantum number is added by the light helicity that carries $\Delta m = -1$ ($\sigma^-$) or $+1$ ($\sigma^+$). The proximity in energy between this Floquet-Bloch band and the equilibrium CB can induce a hybridization that leads to an energy shift provided that they have the same magnetic quantum numbers. Although this material is known to possess a strong excitonic interaction, the physical description of the valley-selectivity still remains the same and, for the purpose of understanding the effect, the energy of the equilibrium CB is essentially reduced by as much as the exciton binding energy \cite{Combescot90}. This explains the valley-selective energy shift we observed in our experiments. These optical selection rules can be described by equation (2) after replacing $\mathcal{M}_{ab}$ by $\mathcal{M}_v$, which now represents the valley selection rules for different laser polarizations \cite{Combescot90}. Additional experiments (below) investigating the dependence on the light intensity and frequency show that the measured energy shift obeys equation (2) extremely well.

Fig 3a shows a series of $\Delta\alpha$ spectra that grow with increasing pump fluence. It can be shown (in Supplementary) that the integrated $\Delta\alpha$ as a function of energy, namely the spectral weight transfer (SWT), is proportional to the energy shift,
\begin{equation}
\int^{\infty}_{E_0}\Delta\alpha(\omega)d\omega = A \Delta E
\end{equation}
where $A$ is the peak absorbance of the sample. In our analysis, it is sufficient to integrate $\Delta\alpha$ in the range of $2.00 \leq \omega \leq 2.18$ eV because the signal vanishes beyond this upper limit. The SWT is plotted in Fig 3b (blue circles) as a function of pump fluence, together with accompanying results measured with smaller (black) and larger (red) laser detuning energies $\Delta$. Not only they show a linear dependence with fluence but they also share a common slope when plotted as a function of fluence$/\Delta$ in Fig 3c. This is in excellent agreement with equation (2). By obtaining the peak absorbance $A = 0.2$ from Fig 1c, we can set an energy scale for $\Delta E$ in the right vertical axis of Fig 3c, which estimates an energy shift of 18 meV measured at the highest fluence. We note that, for a given fluence and energy detuning, this material exhibits the largest optical Stark shift in any materials reported to date.

\begin{figure}[t]
	\includegraphics[width=0.45\textwidth]{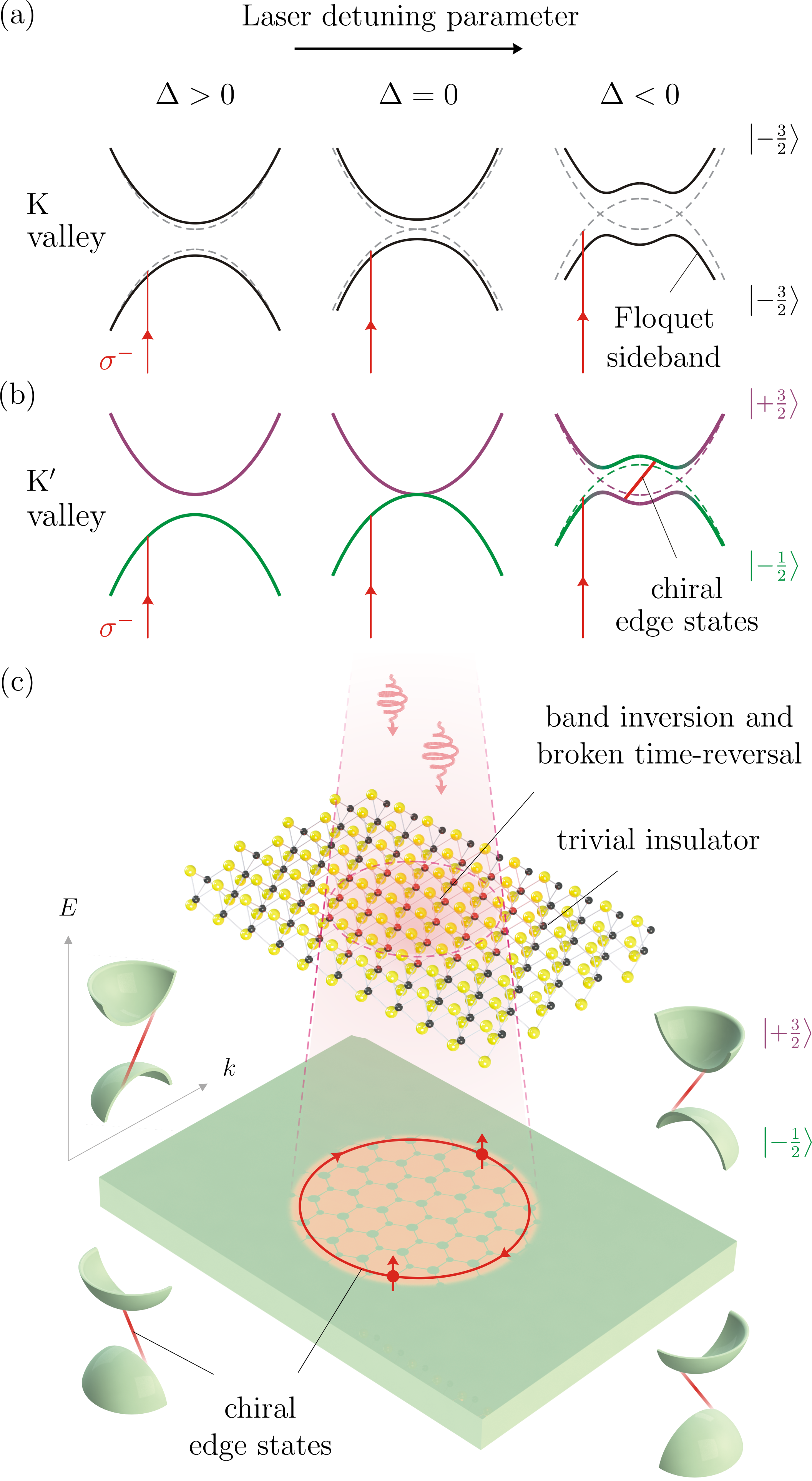}
	\caption{\textbf{Valley-specific Floquet topological phase.} (a,b) Hybridization between the Floquet-Bloch band and the conduction band in variation of $\Delta$, which gives rise to the avoided crossing at K valley (a) and the band inversion at K$^{\prime}$ valley (b). The crossing straight line (red) at K$^{\prime}$ valley is the anticipated chiral edge states due to the band inversion. (c) Schematic of the Floquet-driven chiral edge state along the boundary of the laser-exposed region.}
	\label{fig:Fig4}
\end{figure}

The ability to create valley-specific Floquet-Bloch bands in monolayer TMDs offers a means to induce new topological phases \cite{Inoue10,Kitagawa11,Lindner11} with valley-specific edge states. To elucidate its valley-specificity, we can ignore the excitonic interaction without loss of generality, and we conceive a situation where the $\sigma^-$ detuning energy is varied across the interband transition, as shown in Fig 4a (K valley) and 4b (K$^{\prime}$ valley). In these figures, the coherent absorption of light (red line) from the valence band creates a Floquet-Bloch band close to the conduction band edge (dashed curves) and induces hybridization that results in energy repulsion (solid curves). When the pump detuning is set at $\Delta < 0$, band inversion should occur at K$^{\prime}$ valley which is accompanied by the creation of topological edge states (red) \cite{Bernevig06}. Here the pump helicity breaks time-reversal symmetry, inducing chiral edge states along the boundary of the laser-exposed region where the topological order changes (Fig 4c) \cite{Piskunow14}. In future works, it should be possible to investigate such Floquet chiral edge states provided that the probing signal is made insensitive to contributions from photoexcited excitons. This opens an exciting avenue, which merges Floquet-driven topological transition and valley physics together.

The observation of a large valley-selective optical Stark effect in monolayer WS$_2$ represents the first clear demonstration of broken valley degeneracy in a monolayer TMD. This is possible because of the formation of Floquet-Bloch states via the application of circularly polarized light, which breaks time-reversal symmetry and allows for the lifting of the intervalley spin degeneracy. These findings offer additional control for valley-switching applications on a femtosecond timescale, as well as provide a means to generate new valley-selective topological phases in monolayer TMDs. 

\vspace{0.5cm}
\noindent
\textbf{Methods}
\newline
\small{The sample consists of high-quality monolayers of WS$_2$ that were CVD-grown on sapphire substrates \cite{YHLee12, YHLee13}, and all measurements in this study were conducted at ambient condition (300 K, 1 atm). In our experiments, we used a Ti:sapphire amplifier producing laser pulses with duration of 50 fs. Each pulse was split into two arms. For the pump arm, the pulses were sent to an optical parametric amplifier to generate tunable photon energy below the absorption peak ($h\nu < E_0$), while for the probe arm the pulses were sent through a delay stage and a white-light continuum generator ($h\nu =$ 1.78$-$2.48 eV, chirp-corrected). The two beams were focused at the sample, and the probe beam was reflected to a monochromator and a photodiode for lock-in detection \cite{EJSie13}. By scanning the grating and the delay stage, we were able to measure $\Delta R/R$ (and hence $\Delta\alpha$, \cite{EJSie13}) as a function of energy and time delay. Here, $\Delta\alpha(\omega,t) = \alpha(\omega,t)-\alpha_0 (\omega)$. The pump and probe polarizations were varied separately by two sets of polarizers and quarter-wave plates, allowing us to perform valley-selective measurements, and an additional half-wave plate for tuning the pump fluence (Fig 1d).}

\vspace{0.5cm}
\noindent
\textbf{Acknowledgements}
\newline
\small{The authors acknowledge technical assistance by Qiong Ma and Yaqing Bie during the measurement of the equilibrium absorption of monolayer WS$_2$, and helpful discussions with Zhanybek Alpichshev, Inna M. Vishik, and Yihua Wang. This work is supported by U.S. Department of Energy (DOE) award numbers DE-FG02-08ER46521 and DE-SC0006423 (data acquisition and analysis). Y.-H.L. and J.K. acknowledges support from NSF DMR 0845358 (material growth and characterization). Y.-H.L. also acknowledges partial support from the Ministry of Science and Technology of the Republic of China (103-2112-M-007-001-MY3).}


\begin{thebibliography}{}

\bibitem{DiXiao12}	D. Xiao \textit{et al.}, Phys. Rev. Lett. 108, 196802 (2012).
\bibitem{KFMak12}		K. F. Mak \textit{et al.}, Nature Nano. 7, 494 (2012).
\bibitem{HZeng12}		H. Zeng \textit{et al.}, Nature Nano. 7, 490 (2012).
\bibitem{TCao12}		T. Cao \textit{et al.}, Nature Comms. 3, 887 (2012).
\bibitem{Inoue10}		J. I. Inoue \textit{et al.}, Phys. Rev. Lett. 105, 017401 (2010).
\bibitem{Kitagawa11}	T. Kitagawa, \textit{et al.}, Phys. Rev. B 84, 235108 (2011).
\bibitem{Lindner11}	N. H. Lindner \textit{et al.}, Nature Phys. 7, 490 (2011).
\bibitem{Shirley65}	J. H. Shirley, Phys. Rev. 138, B979 (1965).
\bibitem{Autler55}	S. H. Autler \textit{et al.}, Phys. Rev. 100, 703 (1955).
\bibitem{Bakos77}		J. S. Bakos, Phys. Rep. 31, 209 (1977).
\bibitem{Cohen90}		C. N. Cohen-Tannoudji \textit{et al.}, Phys. Today 43, 33 (1990).
\bibitem{YHWang13}	Y. H. Wang \textit{et al.}, Science 342, 453 (2013).
\bibitem{Joffre88}	M. Joffre \textit{et al.}, J. Mod. Opt. 35, 1951 (1988).
\bibitem{Frohlich85}	D. Fr\"{o}hlich \textit{et al.}, Phys. Rev. Lett. 55, 1335 (1985).
\bibitem{Mysyrowicz86}	A. Mysyrowicz \textit{et al.}, Phys. Rev. Lett. 56, 2748 (1986).
\bibitem{Chemla89}	D. S. Chemla \textit{et al.}, J. Lumin. 44, 233 (1989).
\bibitem{Sieh99}		C. Sieh \textit{et al.}, Phys. Rev. Lett. 82, 3112 (1999).		
\bibitem{Hayat12}		A. Hayat \textit{et al.}, Phys. Rev. Lett. 109, 033605 (2012).
\bibitem{Koster11}	N. S. Koster \textit{et al.}, Appl. Phys. Lett. 98, 161103 (2011).
\bibitem{EJSie13}		E. J. Sie \textit{et al.}, arXiv:1312.2918.
\bibitem{Combescot90}	M. Combescot, Phys. Rev. B 41, 3517 (1990).
\bibitem{Bernevig06}	B. Andrei Bernevig \textit{et al.}, Science 314, 1757 (2006).
\bibitem{Piskunow14} 	P. M. Perez-Piskunow \textit{et al.}, Phys. Rev. B 89, 121401 (2004).
\bibitem{YHLee12} 	Y. -H. Lee \textit{et al.}, Adv. Mat. 24, 2320 (2012).
\bibitem{YHLee13}		Y. -H. Lee \textit{et al.}, Nano Lett. 13, 1852 (2013).

\end{thebibliography}
\end{document}